\documentclass[11pt,letter]{IEEEtran}
\usepackage{graphicx,color}
\usepackage{amsmath,amssymb}
\usepackage{fancyhdr}

\newcommand{\Hi} {{\cal H}^\infty}
\newcommand{\ssp}{{\cal S}_\infty(P)}
\newcommand{\stp}{{\cal S}(P)}
\newcommand{\Dq}{{\cal D}_q}
\newcommand{\bea}{\begin{eqnarray}}
\newcommand{\eea}{\end{eqnarray}}
\newcommand{\be}{\begin{equation}}
\newcommand{\ee}{\end{equation}}
\newcommand{\bd}{\begin{displaymath}}
\newcommand{\ed}{\end{displaymath}}
\newcommand{\rhp}{\mathbb{C}_{+}}
\newcommand{\rhps}{\mathbb{C}_{+}^{\sigma_o}}
\newcommand{\R}{\mathbb{R}}
\newcommand{\ud}{\mathbb{D}}

\begin{document}

\title{Sensitivity Minimization by Strongly Stabilizing Controllers for a Class of Unstable Time-Delay Systems$^*$}

\author{\thanks{$^*$ This work was supported in part by
T\"UB\.{I}TAK (grant no. EEEAG-105E156).} Suat Gumussoy$^\dag$
\thanks{$^\dag$ S.~Gumussoy is with The MathWorks Inc., Natick, MA,
01760, {\sl suat.gumussoy@mathworks.com}} and Hitay
\"Ozbay$^\ddag$ \thanks{$^\ddag$ H.~\"Ozbay is with Dept. of Electrical and
Electronics Eng., Bilkent University, Ankara TR-06800, Turkey,
{\sl hitay@bilkent.edu.tr}}}

\maketitle

\begin{abstract}
Weighted sensitivity minimization is studied within the framework
of strongly stabilizing (stable) $\Hi$~controller design for a
class of infinite dimensional systems.   This problem has been
solved by Ganesh and Pearson, \cite{GP86}, for finite dimensional
plants using Nevanlinna-Pick interpolation. We extend their
technique to a class of unstable time delay systems. Moreover, we
illustrate suboptimal solutions, and their robust implementation.
\end{abstract}
\IEEEpeerreviewmaketitle

\section{Introduction}

In this note the sensitivity minimization problem for a class of
infinite dimensional systems is studied. The goal is to minimize the
$\Hi$ norm of the weighted sensitivity by using stable controllers
from the set of all stabilizing controllers for the given plant.
This problem is a special case  of strongly stabilizing (i.e.
stable) controller design studied earlier, see for example
\cite{CZ01,CZ03,CC01,CWL03,GO05,IOS93,JJA90,LS02,P06,P06cdc,SGP97,SS85,V85,ZO99,ZO00},
and their references for different versions of the problem. The
methods in \cite{B96,GP86} give optimal (sensitivity minimizing)
stable $\Hi$ controllers for finite dimensional SISO plants. Other
methods provide sufficient conditions to find stable suboptimal
$\Hi$ controllers. As far as infinite dimensional systems are
concerned, \cite{GO04,S91} considered systems with time delays.

In this paper, the method of \cite{GP86} is generalized for a
class of time-delay systems. The plants we consider may have
infinitely many right half plane poles. Optimal and suboptimal
stable $\Hi$ controllers are obtained for the weighted sensitivity
minimization problem using the Nevanlinna-Pick interpolation.

It has been observed that (see e.g. \cite{GP86,GO06}) the
Nevanlinna-Pick interpolation approach used in these papers lead
to stable controllers with ``essential singularity'' at infinity.
This means that the controller is non-causal, i.e. it contains a
time advance, as seen in the examples. In this note, by putting a
norm bound condition on the inverse of the weighted sensitivity we
obtain causal suboptimal controllers using the same interpolation
approach. This extra condition also gives an upper bound on the
$\Hi$ norm of the stable controller to be designed. Another method
for causal suboptimal controller design is a rational proper function search in the set of all suboptimal interpolating functions. This method is also illustrated with an example.

The problem studied in the paper is defined in
Section~\ref{sec:2}. Construction procedure for optimal strongly
stabilizing $\Hi$ controller is given in Section~\ref{sec:3}.
Derivation of causal suboptimal controllers is presented in
Section~\ref{sec:4}.  In Section~\ref{sec:example} we give an
example illustrating the methods proposed here for unstable time delay systems. Concluding remarks are made in Section~\ref{sec:6}.

\section{Problem Definition} \label{sec:2}
\begin{figure}[h]
\begin{center}
\includegraphics[width=9cm]{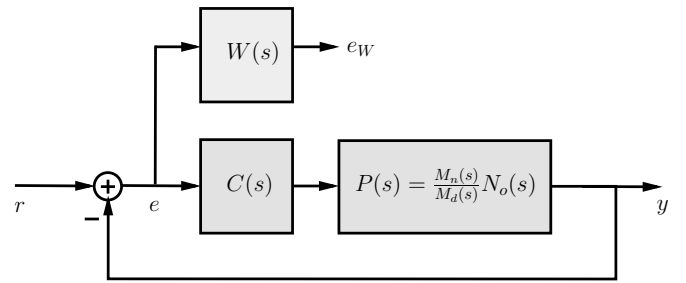}
\caption{\label{fig:standardfeedback} Standard Feedback System}
\end{center}
\end{figure}

Consider the standard unity feedback system with
single-input-single-output plant $P$ and controller $C$ in Figure \ref{fig:standardfeedback}. The
sensitivity function for this feedback system is $S=(1+PC)^{-1}$.
We say that the controller stabilizes the plant if $S$, $CS$ and
$PS$ are in $\Hi$. The set of all stabilizing controllers for a
given plant $P$ is denoted by $\stp$, and we define $\ssp=\stp
\cap \Hi$ as the set of all strongly stabilizing controllers.

For a given minimum phase filter $W(s)$ the classical weighted
sensitivity minimization problem (WSM) is to find \be \gamma_{\rm
o}= \sup_{\substack{r\in\mathcal{L}_2 \\ r\neq0}}
\frac{\|e_W\|_2}{\|r\|_2} =\inf_{C\in\stp} \|W(1+PC)^{-1} \|_\infty
. \label{WSMopt} \ee When we restrict the controller to the set
$\ssp$ we have the problem of weighted sensitivity minimization by a
stable controller (WSMSC): in this case the goal is to find \be
\gamma_{\rm ss}= \inf_{C\in\ssp} \|W(1+PC)^{-1} \|_\infty, \label{WSMssopt} \ee and
the optimal controller $C_{\rm ss,opt}\in\ssp$.

Transfer functions of the plants to be considered here are in the form
\be P(s)=
\frac{M_n(s)}{M_d(s)}N_o(s) \label{eq:plant} \ee where $M_n$, $M_d$
are inner and $N_o$ is outer. We will assume that $M_n$ is rational
(finite Blaschke product), but $M_d$ and $N_o$ can be infinite
dimensional.  The relative degree of $N_o$ is assumed to be an
integer $n_o \in \mathbb{N}$, i.e., we consider plants for which the
decay rate of $20\log(|N_o(j\omega )|)$, as $\omega \rightarrow
\infty$, is $-20~ n_o$ dB per decade, for some non-negative integer
$n_o$.

A typical example of such plants is
retarded or neutral time delay system written in the form
\be \label{eq:delayplant}
P(s)=\frac{R(s)}{T(s)}=\frac{\sum_{i=1}^{n_r} R_i(s)e^{-h_i
s}}{\sum_{j=1}^{n_t} T_j(s)e^{-\tau_j s}}
\ee
where
\begin{enumerate}
\item[(i)] $R_i$ and $T_j$ are stable, proper, finite dimensional
transfer functions, for $i=1,\dots,n_r$, and $j=1,\dots,n_t$;
\item[(ii)] $R$ and $T$ have no imaginary axis zeros, but they may
have finitely many zeros in $\rhp$; moreover, $T$ is allowed to
have infinitely many zeros in $\rhp$, see below cases (ii.a) and (ii.b);
\item[(iii)] time delays,
$h_i$ and $\tau_j$ are rational numbers such that $0=
h_1<h_2<\ldots<h_{n_r}$, and $0=\tau_1<\tau_2<\ldots<\tau_{n_t}$.
\end{enumerate}

In \cite{GO-IFAC-06} it has been shown that under the conditions
given above the time delay system \eqref{eq:delayplant} can be put
into general form \eqref{eq:plant}. In order to do this,
define the conjugate of $T(s)$ as
$\bar{T}(s):=e^{-\tau_{n_t}s}T(-s)M_C(s)$ where $M_C$ is inner,
finite dimensional whose poles are poles of $T$.
For notational convenience, we say that $T$ is an $F$-system
(respectively, $I$-system) if $T$ (respectively,
$\bar{T}$) has finitely  many zeros in $\rhp$; (note that when $T$ is an $I$-system the plant has infinitely many poles in $\rhp$).
The plant
factorization can be done as follows for two different cases:
\newline
\hspace*{1cm}\underline{Case (ii.a)}: When $R$ is an $F$-system
and $T$ is an $I$-system:
\be  \label{eq:FIfact}
M_n=M_R,\quad M_d=M_{\bar{T}}\frac{T}{\bar{T}}, \quad
N_o=\frac{R}{M_R}\frac{M_{\bar{T}}}{\bar{T}},
\ee
\hspace*{1cm}\underline{ Case (ii.b)}: When $R$ and $T$ are both
$F$-systems:
\be \label{eq:FFfact}
M_n=M_R,\quad M_d=M_{T},\quad N_o=\frac{R}{M_R}\frac{M_{T}}{T}
\ee
The inner functions, $M_R$, $M_T$ and $M_{\bar{T}}$, are defined
in such a way that their zeros are $\rhp$ zeros of $R$, $T$ and
$\bar{T}$, respectively. By assumption (ii), $R$, $T$ (case
(ii.b)) and $\bar{T}$ (case (ii.a)) have finitely many zeros in
$\rhp$, so, the inner functions, $M_R$, $M_T$ and $M_{\bar{T}}$
are finite dimensional.


\noindent {\bf Example}. Consider a plant with infinitely many
poles in $\rhp$ (this corresponds to case (ii.a) where $R$ and $T$
are $F$-system and $I$-system respectively; clearly, the plant
factorization in case (ii.b) is much easier):
\begin{eqnarray}
P_{FI}(s)&=&\frac{(s+1)+4e^{-3s}}{(s+1)+2(s-1)e^{-2s}}
\label{plant:FIplant} \\
&=&\frac{R(s)}{T(s)}=\frac{1e^{-0s}+
\left(\frac{4}{s+1}\right)e^{-3s}}{1e^{-0s}+
\left(\frac{2s-2}{s+1}\right)e^{-2s}}. \nonumber
\end{eqnarray}
It can be shown that $R$ has only two $\rhp$ zeros at
$s_{{1,2}}\approx 0.3125 \pm j0.8548$. Also, $T$ has infinitely
many $\rhp$ zeros converging to $\ln{\sqrt{2}}\pm
j(k+\frac{1}{2})\pi$ as $k\rightarrow\infty$. In this case relative degree is $n_o=0$, and the plant can be
re-written as \eqref{eq:plant} with $\bar{T}(s)=e^{-2s}T(-s)\left(\frac{s-1}{s+1}\right)
=2+\left(\frac{s-1}{s+1}\right)e^{-2s}$, 
\bea \label{factorizedplant}
M_n(s)&=&\frac{(s-s_{1})(s-s_{2})}
                        {(s+s_{1})(s+s_{2})},\quad
                             M_d(s)=\frac{T(s)}{\bar{T}(s)}, \\
 \nonumber
 N_o(s)&=&\frac{R(s)}{M_n(s)}\frac{1}{\bar{T}(s)}
  .
 \eea

\section{Optimal Weighted Sensitivity}\label{sec:3}

In this section we illustrate how the Nevanlinna-Pick approach
proposed in \cite{GP86} extends to the classes of plants in the
form \eqref{eq:delayplant}. We will also see that the optimal
solution in this approach leads to a non-causal optimal
controller. In the next section we will modify the interpolation
problem to solve this problem.

First, in order to eliminate a technical issue, which is not
essential in the weighted sensitivity minimization, we will
replace the outer part, $N_o$, of the plant with
\bd
N_\varepsilon(s)=N_o(s)(1+\varepsilon s)^{n_o}
\ed
where $\varepsilon >0$ and $\varepsilon \rightarrow 0$. This makes
sure that the plant does not have a zero at $+\infty$, and hence
we do not have to deal with interpolation conditions at infinity.
See \cite{F92,FZ84} for more discussion on this issue and justification of approximate inversion of the outer part of the plant in weighted sensitivity minimization problems.

Now, let $s_1,\dots,s_n$ be the zeros of $M_n(s)$ in $\rhp$. Then,
WSMSC problem can be solved by finding a function $F(s)$
satisfying three conditions (see e.g. \cite{DFT92,GP86,V85})
\newline
\hspace*{1cm} (F1) $F\in\Hi$ and $\| F\|_\infty \le 1$;\newline
\hspace*{1cm} (F2) $F$ satisfies interpolation conditions
\eqref{F:interp}; \newline \hspace*{1cm} (F3) $F$ is a unit in
$\Hi$, i.e. $F, F^{-1}\in\Hi$;
\be
F(s_i) = \frac{W(s_i)}{\gamma M_d(s_i)}
=:\frac{\omega_i}{\gamma},\quad i=1,\ldots,n. \label{F:interp}
\ee

Once such an $F$ is constructed, the controller
\be \label{eq:Cgamma}
C_{\gamma}(s)=\frac{W(s)-\gamma M_d(s) F(s)}{\gamma M_n(s) F(s) }
~N_\varepsilon (s)^{-1}
\ee
is in $\ssp$ and it leads to $\|W(1+PC)^{-1} \|_\infty \le
\gamma$. Therefore, $\gamma_{\rm ss}$ is the smallest $\gamma$ for
which there exists $F(s)$ satisfying F1, F2 and F3. It is also
important to note that the controller \eqref{eq:Cgamma} is the
solution of the unrestricted weighted sensitivity minimization
(WSM) problem, defined by \eqref{WSMopt}, when $F(s)$ satisfies F1
and F2 for the smallest possible $\gamma>0$; in this case, since
F3 may be be violated, the controller may be unstable.

The problem of constructing $F(s)$ satisfying F1--F3 has been
solved by using the Nevanlinna-Pick interpolation as follows.
First define
\be
\label{def:G(s)} G(s)=-\ln{F(s)} ~,~~~F(s)=e^{-G(s)}.
\ee
Now, we want to find an analytic function $G~:~\rhp \rightarrow
\rhp$ such that
\be
G(s_i)=-\ln{\omega_i}+\ln{\gamma}-j2\pi \ell_i
       =:\nu_i,\quad i=1,\ldots,n \label{nu_i}
\ee
where $\ell_i$ is a free integer due to non-unique phase of the
complex logarithm. Note that when $\|F\|_\infty \le 1$ the
function $G$ has a positive real part hence it maps $\rhp$ into
$\rhp$. Let $\mathbb{D}$ denotes the open unit disc, and transform
the problem data from $\rhp$ to $\ud$ by using a one-to-one
conformal map $z=\phi(s)$. The transformed interpolation
conditions are
\be
\label{eq:interpcondz} f(z_i)=\frac{\omega_i}{\gamma},\quad
i=1,\ldots,n
\ee
where $z_i=\phi(s_i)$ and $f(z)=F(\phi^{-1}(z))$. The transformed
interpolation problem is to find a unit with $\|f\|_\infty\leq 1$
such that interpolation conditions (\ref{eq:interpcondz}) are
satisfied. By the  transformation $g(z)=-\ln{f(z)}$, the
interpolation problem can be written as,
\be
g(z_i)=\nu_i,\quad i=1,\ldots,n.
\ee
Define $\phi(\nu_i)=:\zeta_i$. If we can find an analytic function
$\tilde{g}~:~\ud \rightarrow \ud$ , satisfying
\be
\tilde{g}(z_i)=\zeta_i ~~~~i=1,\ldots,n \label{eq:gtilde}
\ee
then the desired $g(z)$, hence $f(z)$ and $F(s)$ can be
constructed from $g(z)=\phi^{-1}(\tilde{g}(z))$. The problem of
finding such $\tilde{g}$ is the well-known Nevanlinna-Pick
problem, \cite{FOT96,KN77,ZO_NevPick}. The condition for the
existence of an appropriate $g$ can be given directly: there
exists such an analytic function $g~:~\ud \rightarrow \rhp$ if and
only if the Pick matrix ${\mathcal P}$,
\be
\label{eq:pickmatrix} \mathcal{P}(\gamma,\{\ell_i,\ell_k\})_{i,k}
=\left[\frac{2\ln{\gamma}-\ln{\omega_i} -\ln{\bar{\omega}_k}+j2\pi
\ell_{k,i}}{1-z_i\bar{z}_k}\right]
\ee
is positive semi-definite, where $\ell_{k,i}=\ell_k-\ell_i$ are
free integers. In \cite{GP86}, it is mentioned that the possible
integer sets $\{\ell_i,\ell_k\}$ are finite and there exists a
minimum value, $\gamma_{\rm ss}$, such that
$\mathcal{P}(\gamma_{\rm ss}, \{\ell_i,\ell_k\})\geq~0$.

The Nevanlinna-Pick problem posed above can be solved as outlined
in \cite{FOT96,KN77,ZO_NevPick}. As noted in \cite{GP86,GO06} and
we illustrate with an example in Section~\ref{sec:example}, generally, as $\gamma$ decreases to $\gamma_{\rm ss}$ the function
$G(s)$ satisfies
\bd
 G(s)\rightarrow k_\gamma s ~,~~{\rm where}~
~~k_\gamma\in\R_+ ~{\rm as}~~s\rightarrow \infty.
\ed
Therefore, in the optimal case $F(s)$ has an essential singularity
at infinity, i.e.,
$\lim_{s\rightarrow \infty}|F(s)|=0$, thus $F^{-1}$ is not bounded in $\rhp$, i.e., $F^{-1}\notin\Hi$.
Clearly, this violates one of the design
conditions and leads to a non-causal controller \eqref{eq:Cgamma},
which typically contains a time advance. In the next section to circumvent this problem we propose to put an $\Hi $ norm bound on $F^{-1}$.

Suboptimal solution of weighted sensitivity minimization (\ref{WSMssopt}) by stable controller
 is similar to the optimal case. The suboptimal controller can be represented as in (\ref{eq:Cgamma}) where
 $\gamma>\gamma_{ss}$. The controller synthesis problem can be reduced into calculation of interpolation
 function $F(s)$ satisfying the conditions F$1$, F$2$ and F$3$. By similar approach used in optimal case,
 the conditions are satisfied if $\tilde{g}$ is calculated satisfying the interpolation conditions (\ref{eq:gtilde}).
 This  is well-known suboptimal Nevanlinna-Pick problem and the parametrization of the solution for suboptimal case
 is given in \cite{FOT96}. After the parametrization is calculated, the controller parametrization (\ref{eq:Cgamma})
 can be obtained by back-transformations as explained above.

\section{Modified Interpolation Problem} \label{sec:4}

The controller \eqref{eq:Cgamma} gives the following weighted
sensitivity
\be \label{eq:sens}
W(s)(1+P(s)C_\gamma(s))^{-1}=\gamma M_d(s)F(s)
\ee
where $F,F^{-1}\in\Hi$, $\|F\|_\infty \le 1$ and \eqref{F:interp}
hold. Since one of the conditions on $F$ is to have $F^{-1}\in
\Hi$ it is natural to consider a norm bound
\be \label{Finv_norm}
\| F^{-1}\|_\infty \le \rho
\ee
for some fixed $\rho>1$. This also puts a bound on the $\Hi$ norm
of the  controller; more precisely,
\be
\| C_\gamma\|_\infty \le \| N_o\|_\infty^{-1} \left( 1
+\frac{\rho}{\gamma}\| W\|_\infty\right) .
\ee

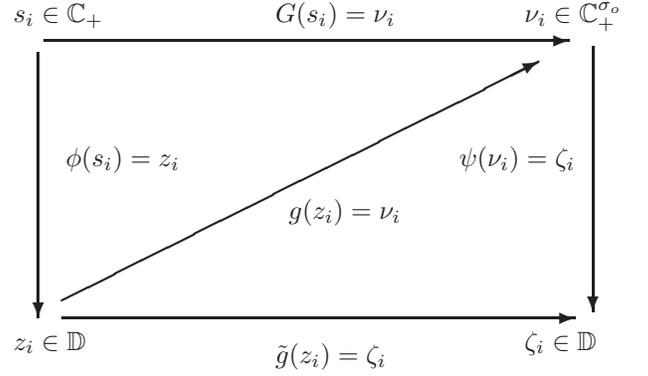
\begin{figure}[ht]
\begin{small}
\setlength{\unitlength}{3250sp}%
\begin{picture}(5000,3000)(2116,-3737)
\thicklines {\put(2836,-1141){\vector( 1, 0){4005}}
}%
{\put(2791,-1231){\vector( 0,-1){2025}}
}%
{\put(2971,-3256){\vector( 1, 0){3915}}
}%
{\put(7021,-1186){\vector( 0,-1){2025}}
}%
{\put(2971,-3121){\vector( 2, 1){3636}}
}%
\put(6500,-1000){$\nu_i\in \mathbb{C}_+^{\sigma_o}$}%
\put(2600,-1000){$s_i\in\mathbb{C}_+$}%
\put(2600,-3500){$z_i\in \mathbb{D}$}%
\put(6500,-3500){$\zeta_i\in \mathbb{D}$}%
\put(3000,-2100){$\phi(s_i)=z_i$}%
\put(6000,-2100){$\psi(\nu_i)=\zeta_i$}%
\put(4600,-1000){$G(s_i)=\nu_i$}%
\put(4600,-3600){$\tilde{g}(z_i)=\zeta_i$}%
\put(4700,-2500){$g(z_i)=\nu_i$}%
\end{picture}%
\caption{Interpolating functions and conformal maps}
\label{conformal.fig}
\end{small}
\end{figure}

Recall that we are looking for an $F$ in the form $F(s)=e^{-G(s)}$,
for some analytic $G:\rhp \rightarrow \rhp$ satisfying
$G(s_i)=\nu_i$, $i=1,\dots ,n$. In this case we will have
$|F(s)|=|e^{-{\rm Re}(G(s))}|\le 1$ for all $s\in \rhp$. On the
other hand, $F^{-1}(s)=e^{G(s)}$. Thus, in order to satisfy
\eqref{Finv_norm}, $G$ should have a bounded real part, namely \be
0<{\rm Re}(G(s))<\ln(\rho)=:\sigma_o \ee Accordingly, define
$\rhps:=\{ s\in \rhp :  0< {\rm Re}(s) <\sigma_o\}$. Then, the
analytic function $G$ we construct should take $\rhp$ into $\rhps$.
Note from \eqref{nu_i} that in order for this modified problem to
make sense $\gamma$ and  $\rho$ should satisfy the following
inequality so that we have a feasible interpolation data, i.e.
$\nu_i\in \rhps$,
\be
\max{\{|\omega_1|,\ldots,|\omega_n|\}}<\gamma<\rho+\max{\{|\omega_1|,
\ldots,|\omega_n|\}}.
\ee
Now take a conformal map $\psi:\rhps \rightarrow \ud$, and set
$\zeta_i:=\psi(\nu_i)$, $z_i=\phi(s_i)$, where as before $\phi$ is a
conformal map from $\rhp$ to $\ud$. Then, the problem is again
transformed to a Nevanlinna-Pick interpolation: find an analytic
function $\tilde{g}:\ud \rightarrow\ud$ such that
$\tilde{g}(z_i)=\zeta_i$, $i=1,\dots n$. Once $\tilde{g}$ is
obtained, the function $G$ is determined as
$G(s)=\psi^{-1}(\tilde{g}(\phi(s)))$. Typically, we take 
$\phi(s)=\frac{s-1}{s+1}$
\begin{eqnarray}
\phi^{-1}(z)&=&\frac{1+z}{1-z}\nonumber \\
\psi(\nu)&=&\frac{je^{-j\pi \nu/\sigma_o}-1}{je^{-j\pi
\nu/\sigma_o}+1} \nonumber \\
\psi^{-1}(\zeta)&=&\frac{\sigma_o}{\pi}\left( \frac{\pi}{2}+ j
\ln (\frac{1+\zeta}{1-\zeta})\right) , \label{conformal:map}
\end{eqnarray}
see e.g. \cite{Nehari_Book}. Interpolating functions defined above
are illustrated by Figure~\ref{conformal.fig}.

It is interesting to note that in this modified problem $\gamma_{\rm
ss}$ (smallest $\gamma$ for which a feasible $\tilde{g}$ exists)
depends on $\rho$, so we write $\gamma_{\rm ss,\rho}$. As $\rho$
decreases, $\gamma_{\rm ss,\rho }$ will increase; and as
$\rho\rightarrow \infty$,  $\gamma_{\rm ss,\rho}$ will converge to
$\gamma_{\rm ss}$, the value found from the unrestricted
interpolation problem  summarized in Section~\ref{sec:3}.

\section{An Example} \label{sec:example}
Consider the plant \eqref{plant:FIplant} defined earlier. Recall
that it has only two $\rhp$ zeros at $s_{{1,2}}\approx 0.3125 \pm
0.8548j$. Let the weighting function be given as
\be
W(s)=\frac{1+0.1s}{s+1}. \label{eq:weightfunction}
\ee
Then, the interpolation conditions are $\omega_{1,2}=0.79\mp
0.42j$. Applying the procedure of \cite{GO06}, summarized in
Section~\ref{sec:3}, we find $\gamma_{\rm ss}=1.0704$. The optimal
interpolating function is
\be
F(s)=e^{-0.57s} \label{Fopt_sol}
\ee
and hence the optimal controller is written as
\be
C_{\gamma_{\rm ss}}=\frac{\frac{1+0.1s}{s+1}-1.0704
\left(\frac{s+1+2(s-1)e^{-2s}}{2(s+1)+(s-1)e^{-2s}}\right)e^{-0.57s}}
{1.0704\left(\frac{s+1+4e^{-3s}}{2(s+1)+(s-1)e^{-2s}}\right)e^{-0.57s}}.
\ee
Clearly, $F^{-1}\notin \Hi$ and the controller is non-causal, it
includes a time advance $e^{+0.57s}$.

\begin{figure}[h]
\begin{center}
\includegraphics[width=8.5cm]{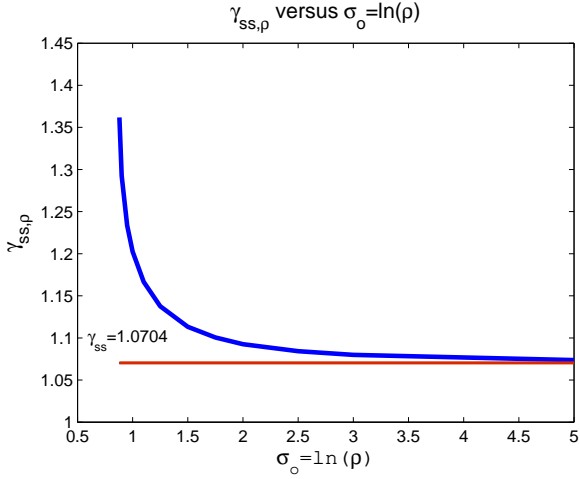}
\end{center}
\caption{\label{Fig_ex} $\gamma_{\rm ss,\rho}$ versus
$\rho=e^{\sigma_o}$}
\end{figure}

If we now apply the modified interpolation idea we see that as
$\rho \rightarrow \infty$ the smallest $\gamma$ for which the
problem is solvable, i.e. $\gamma_{\rm ss}$, approaches to 1.0704,
which is the optimal performance level found earlier. On the other
hand, as $\rho $ decreases $\gamma_{\rm ss}$ increases, and there
is a minimum value of $\rho=e^{0.88}=2.41$, below which there is
no solution to the interpolation problem. See Figure~\ref{Fig_ex}.

\begin{figure}[h]
\begin{minipage}[t]{8cm}
\begin{center}
\includegraphics[width=8cm]{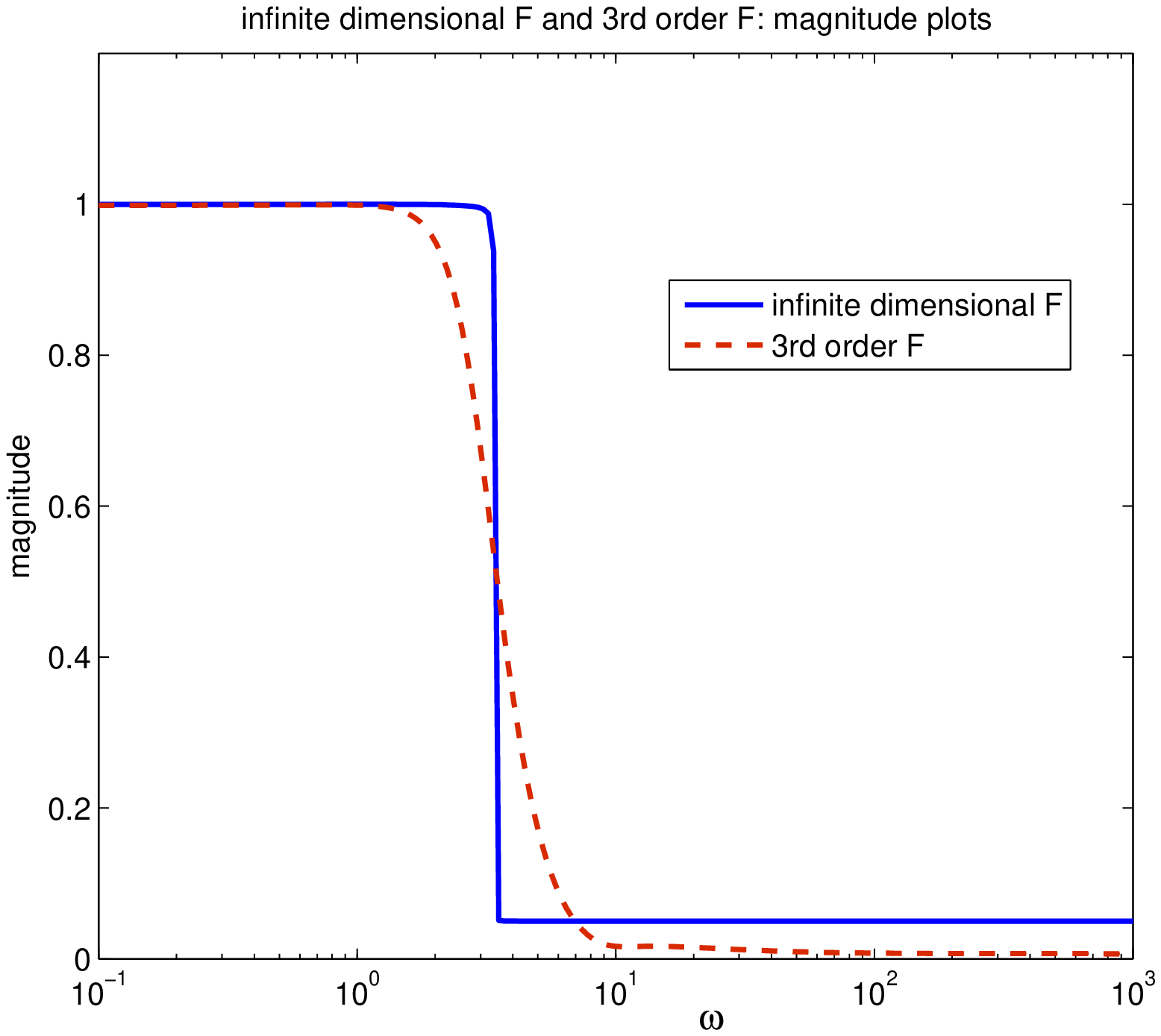}
\end{center}
\end{minipage}
\hfill
\begin{minipage}[t]{8cm}
\begin{center}
\includegraphics[width=8cm]{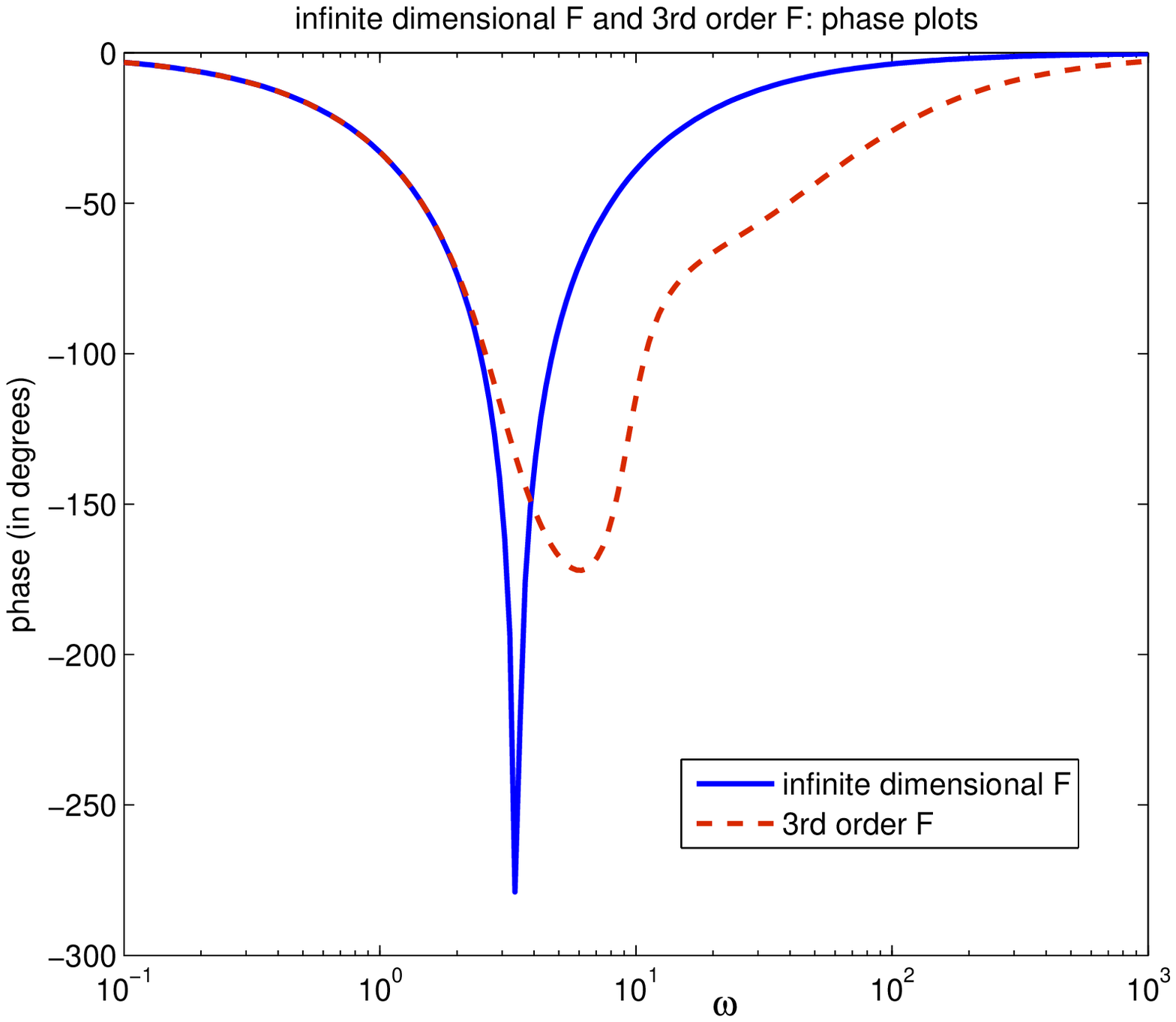}
\end{center}
\end{minipage}
\caption{Magnitude and phase plots of $F$ given in
\eqref{F_opt_ModNP} and \eqref{3rd_order_F}.}\label{F_and_Fapprox}
\end{figure}

For $\sigma_o=3$, i.e. $\rho=e^3\approx 20$, we have $\gamma_{\rm
ss,\rho}=1.08$, and the resulting interpolant is given by
\be
\tilde{G}(s):=\tilde{g}(\phi(s))=j\frac{-0.99794(s-3.415)(s+1)}{
  (s+3.406) (s+1.001)}.
\ee
The optimal $F(s)=e^{-G(s)}$ is determined from
\be
G(s)=\psi^{-1}(\tilde{G}(s))
\ee
where $\psi^{-1}$ is as defined in \eqref{conformal:map}. The
optimal $F$ is
\be
F(s)= \exp(-\frac{\sigma_o}{2}-\frac{j\sigma_o}{\pi}
\ln(\frac{1+\tilde{G}(s)}{1-\tilde{G}(s)})). \label{F_opt_ModNP}
\ee
Note that
the optimal $F(s)$ is infinite dimensional. The magnitude and
phase of $F(j\omega)$ are shown in Figure~\ref{F_and_Fapprox}.
Rational  approximations of \eqref{F_opt_ModNP} can be obtained from the frequency response data using approximation techniques for stable minimum phase infinite dimensional systems, see e.g. \cite{A01}, \cite{GKL89}, \cite{M90}, and their references.

Another way to obtain  finite dimensional interpolating function
$F(s)$ is to search for a proper free parameter in the set of all
suboptimal solutions to the interpolation problem of finding $F$
 satisfying F1--F3.
For a given $\gamma>\gamma_{\rm ss}$ we can parameterize all
suboptimal solutions to this problem as, (see e.g. \cite{FOT96})
\be
\label{eq:interpfunction}
f(z)=\frac{\tilde{P}(z)q(z)+\tilde{Q}(z)}{P(z)+Q(z)q(z)},\quad
\|q\|_\infty\leq 1 ,
\ee
where  $\tilde{P}, \tilde{Q}, P, Q$ are computed as in
\cite{FOT96,KN77,ZO_NevPick}. Using first-order free parameter
\be
q(z)=\frac{az+b}{z+c},
\ee
we search for a unit $f$ in the set determined by
(\ref{eq:interpfunction}). Since $\|q\|_\infty\leq1$, the
parameters $(a,b,c)$ are in the set
\begin{small}
\be
\Dq:=\left\{(a,b,c): |c|\geq1,
\;|a+b|\leq|c+1|,\;|a-b|\leq|c-1|\right\}.
\ee
\end{small}
Then a unit function $f$ can be found if there exist
$(a,b,c)\in~\Dq$ such that
\be \label{eq:charpoly}
(az+b)\tilde{P}(z)+(z+c)\tilde{Q}(z)
\ee
has no zeros in $\ud$. The problem of finding $(a,b,c)$ such that
(\ref{eq:charpoly}) has no zeros in $\ud$ is equivalent to
stabilization of discrete-time systems by first-order controllers
considered in \cite{TKB03}. So we take the intersection of the
parameters found using \cite{TKB03} and the set $\Dq$. The
stabilization set $(a,b,c)$ is determined by fixing $c$ and
obtaining the stabilization set in $a-b$ plane by checking the
stability boundaries.

\begin{figure}[h]
\begin{minipage}[t]{8cm}
\begin{center}
\includegraphics[width=6.5cm]{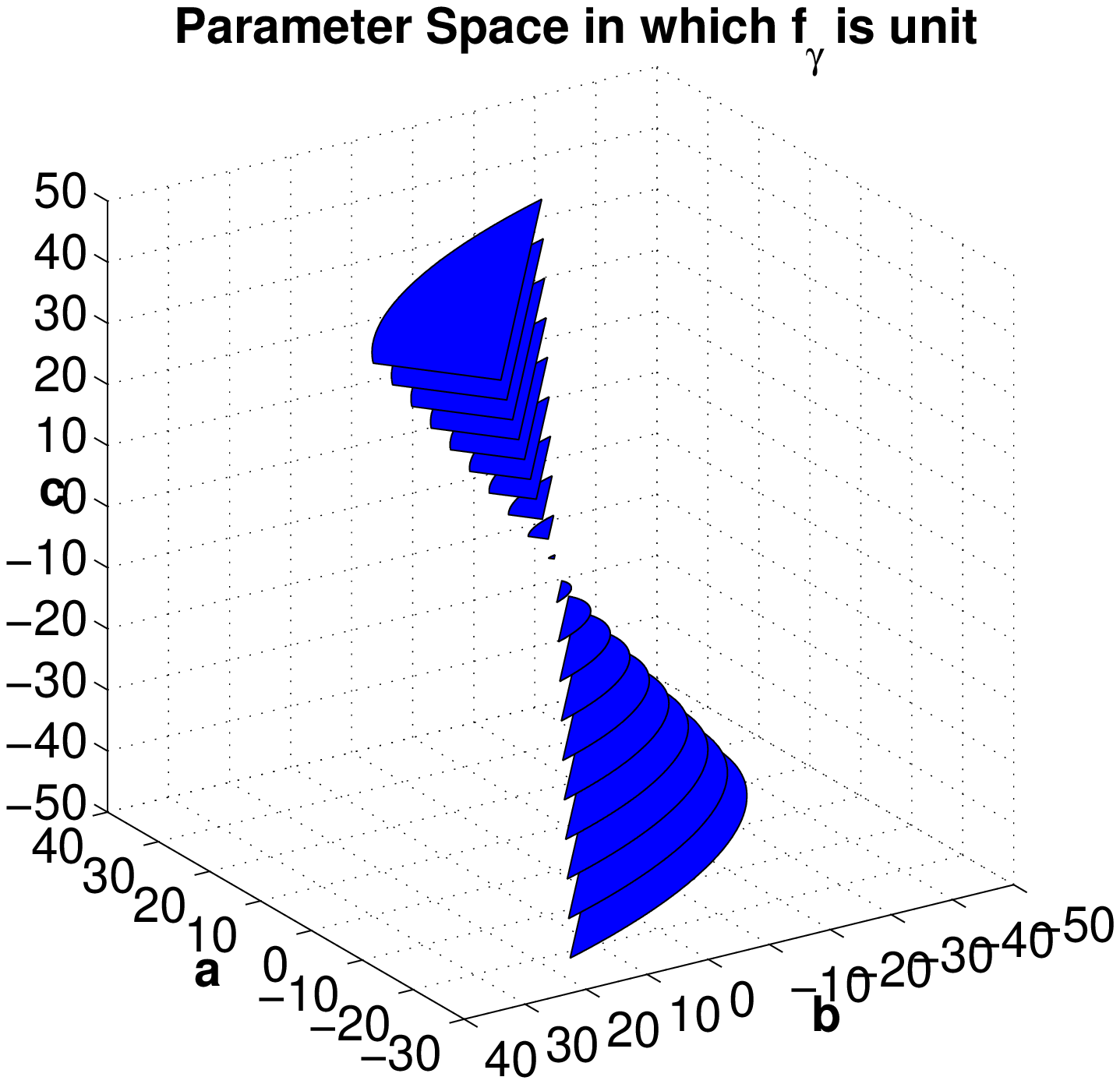}
\caption{\label{Figparamspace} Feasible $(a,b,c)$ for $f$ to be a
unit.}
\end{center}
\end{minipage}
\hfill
\begin{minipage}[t]{8cm}
\begin{center}
\includegraphics[width=6.5cm]{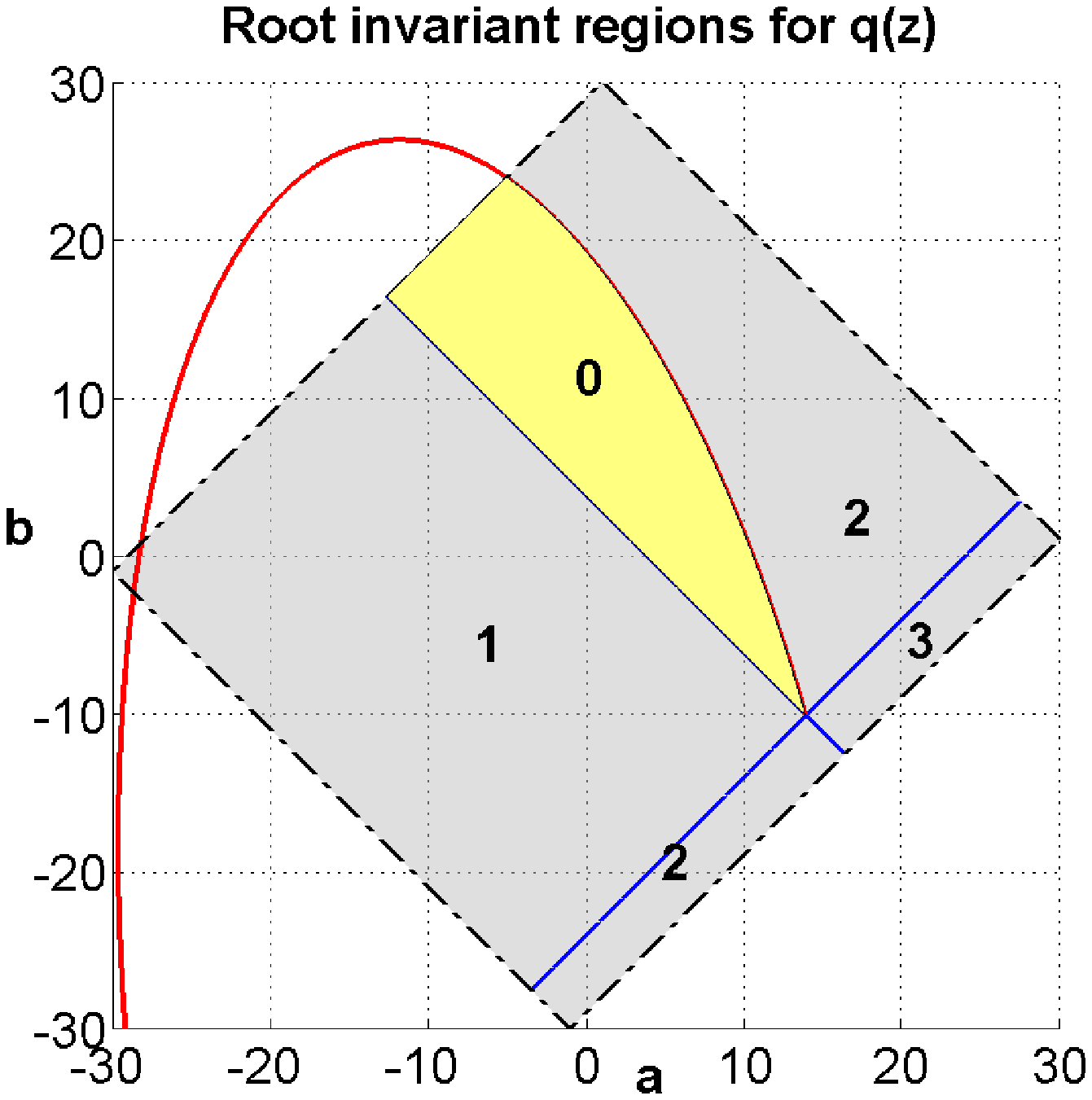}
\caption{\label{Fig_invreg30} Root invariant regions for $c=30$.}
\end{center}
\end{minipage}
\end{figure}

For the above example, let $\gamma=1.2>1.07=\gamma_{\rm ss}$.
After the calculation of $\tilde{P}$, $\tilde{Q}$, $P$, $Q$, we obtain feasible parameter pairs $(a,b)$, for each fixed $c$,
resulting in a unit $f(z)$  as shown in Figure
\ref{Figparamspace}. Note that all values in $(a,b,c)$ parameter
set results in stable suboptimal $\Hi$ controller which gives
flexibility in design to meet other design requirements.

In Figure \ref{Fig_invreg30}, stability region for
(\ref{eq:charpoly}) is given for $c=30$. Red and blue lines are
real and complex-root crossing boundaries respectively. The yellow
colored region (labeled as region {\sf 0} in the grayscale print)
is the area, where the polynomial (\ref{eq:charpoly}) has no $\rhp$ zeros and
the corresponding $\Hi$ controller is stable. The value of
$\gamma=1.2$ is chosen to show the controller parameterization set
and stability regions clearly. If we apply the same technique for
$\gamma=1.08$ the feasible region in $\R^3$ shrinks, but we still
get a solution:
\be
F(s)=\frac{0.068s^3+3.77s^2+21.45s+295.84}{9.93s^3+62.77s^2
+187.25s+296.27}. \label{3rd_order_F}
\ee
It is easy to verify that
\be F(s_i)=\frac{\omega_i}{1.08},~~~~{\rm for}~~i=1,2.
\ee
The function $F$ is a unit with poles and zeros
\begin{eqnarray}
{\tt zero(F)} &=&  -50.9245, -2.2583 \pm j~8.9628\\
{\tt pole(F)} &=&   -3.3510, -1.4851 \pm j~2.5881
\end{eqnarray}
and from its Bode plot we find
$\|F\|_\infty=\frac{295.84}{296.27}<1$. Moreover, $F^{-1}\in \Hi $
with $\| F^{-1}\|_\infty \approx 146$.

In order to compare the third order $F$ given in
\eqref{3rd_order_F}, with the infinite dimensional $F$ described by
\eqref{F_opt_ModNP}, (both of them are designed for $\gamma=1.08$)
we provide their magnitude and phase plots in
Figure~\ref{F_and_Fapprox}.


Although finding a finite dimensional $F(s)$ results in
infinite dimensional suboptimal controller $C_\gamma(s)$, \eqref{eq:Cgamma}, it is possible to implement the controller in a stable manner using the ideas of \cite{GO-IFAC-06} as discussed in early versions of the current paper \cite{GO06,GO07}.

\begin{figure}[h]
\begin{center}
\includegraphics[width=9cm]{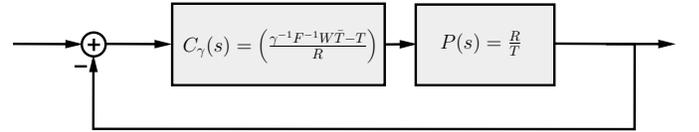}
\caption{\label{fig:closedloopWithController} Feedback System with Controller and Plant Considered in the Example.}
\end{center}
\end{figure}

The structure of the controller for this particular example is in the form
\be
 C_\gamma(s)= \left(\frac{\gamma^{-1}
F^{-1}(s)W(s)\bar{T}(s)-T(s)}{R(s)}\right) ,
\ee
and the overall closed loop system is as shown in Figure~\ref{fig:closedloopWithController}.
Note that at the right half plane zeros of $R(s)$ the numerator vanishes due to interpolation conditions on $F(s)$.
This fact and that $F^{-1}$ is stable makes the controller stable.


Also, one can see that both modified interpolation problem solution with infinite dimensional $F$ (\ref{F_opt_ModNP}) and finite dimensional $F$ (\ref{3rd_order_F}) satisfies sensitivity design constraints. So, the controller is strongly stabilizing (closed loop system is stable with a stable controller), and by (\ref{eq:sens}), the magnitude of weighted sensitivity function on
the imaginary axis is equal to
\be |W(1+PC)^{-1}|=|\gamma
M_d(j\omega)F(j\omega)|=\gamma |F(j\omega)|.
\ee
Therefore, the
magnitude of $F$ on the imaginary axis is equivalent to magnitude of
normalized weighted sensitivity function on the imaginary axis. Both
sensitivity functions satisfies the $\Hi$ norm requirement for all
frequencies. The controllers also achieve good tracking  for low
frequency signals as aimed by selection of weighting function $W$
(\ref{eq:weightfunction}).

\section{Conclusions} \label{sec:6}

In this note we have modified the Nevanlinna-Pick interpolation
problem appearing in the computation of the optimal strongly
stabilizing controller minimizing the weighted sensitivity. By
putting a bound on the norm of $F^{-1}$, a bound on the $\Hi$ norm
of the controller can be obtained.  We have obtained the optimal
$\gamma_{\rm ss,\rho}$ as a function of $\rho$, where $
\|F^{-1}\|_\infty\le \rho$. The example illustrated that as $\rho \rightarrow \infty$,
$\gamma_{\rm ss,\rho}$ converges to the optimal $\gamma_{\rm ss}$
for the problem where $\|F^{-1}\|_\infty$ is not constrained. The
controller obtained here is again infinite dimensional; for
practical purposes it needs to be approximated by a rational
function.  In general this method may require very high order
approximations since the order of strongly stabilizing controllers
for a given plant (even in the finite dimensional case) may have
to be very large, \cite{SS86}. Another method for finding a low
order $F$ satisfying all the conditions is also illustrated with
the given example.  It searches for a first order free parameter
leading to a unit $f$.

\end{document}